\documentclass[letterpaper,twocolumn,prl,aps,superscriptaddress,amsmath,amssymb,floatfix]{revtex4-1}
\usepackage[utf8]{inputenc}
\usepackage[T1]{fontenc}    
\usepackage{color}
\usepackage{amsmath}
\usepackage{amssymb}
\usepackage{graphicx}
\usepackage{esint}
\usepackage[unicode=true,
 bookmarks=true,bookmarksnumbered=false,bookmarksopen=false,
 breaklinks=false,pdfborder={0 0 1},backref=false,colorlinks=true]
 {hyperref}
\hypersetup{
 linkcolor=magenta,urlcolor=blue,citecolor=blue,pdfstartview={FitH},hyperfootnotes=false}

\makeatletter

\usepackage{textcomp}
\usepackage{epstopdf}

\pdfpageheight\paperheight
\pdfpagewidth\paperwidth

\@ifundefined{textcolor}{}{%
 \definecolor{BLACK}{gray}{0}
 \definecolor{WHITE}{gray}{1}
 \definecolor{RED}{rgb}{1,0,0}
 \definecolor{GREEN}{rgb}{0,1,0}
 \definecolor{BLUE}{rgb}{0,0,1}
 \definecolor{CYAN}{cmyk}{1,0,0,0}
 \definecolor{MAGENTA}{cmyk}{0,1,0,0}
 \definecolor{YELLOW}{cmyk}{0,0,1,0}
}

\usepackage{soul}

\usepackage{xcolor}\usepackage{soul}
\def\ket#1{| #1 \rangle}
\def\bra#1{\langle #1 |}

\makeatother

\begin{document}

%\title{Indirectly measuring an observable by coupling with a quantum meter}

\title{Deterministic Generation of Arbitrary Fock States via Resonant Subspace Engineering}

\affiliation{Institute of Fundamental and Frontier Sciences, University of Electronic Science and Technology of China, Chengdu, 610051, China}
\affiliation{Laboratory of Quantum Information,
University of Science and Technology of China, Hefei 230026, China}
\affiliation{Center for Quantum Information, Institute for Interdisciplinary Information Sciences, Tsinghua University, Beijing, China}
\affiliation{Yangtze Delta Industrial Innovation Center of Quantum Science and Technology, Suzhou 215000, China}
\affiliation{CAS Center For Excellence in Quantum Information and Quantum Physics,
University of Science and Technology of China, Hefei, Anhui 230026,
China}
\affiliation{Anhui Province Key Laboratory of Quantum Network,
University of Science and Technology of China, Hefei 230026, China}
\affiliation{Hefei National Laboratory, Hefei 230088, China}

\author{Shan Jin}
\thanks{These authors contribute equally to this work.}
\affiliation{Institute of Fundamental and Frontier Sciences, University of Electronic Science and Technology of China, Chengdu, 610051, China}
\affiliation{Yangtze Delta Industrial Innovation Center of Quantum Science and Technology, Suzhou 215000, China}

\author{Ming Li}
\thanks{These authors contribute equally to this work.}
\affiliation{Laboratory of Quantum Information,
University of Science and Technology of China, Hefei 230026, China}
\affiliation{CAS Center For Excellence in Quantum Information and Quantum Physics,
University of Science and Technology of China, Hefei, Anhui 230026,
China}
\affiliation{Anhui Province Key Laboratory of Quantum Network,
University of Science and Technology of China, Hefei 230026, China}
\affiliation{Hefei National Laboratory, Hefei 230088, China}

\author{Weizhou Cai}
\affiliation{Laboratory of Quantum Information,
University of Science and Technology of China, Hefei 230026, China}
\affiliation{CAS Center For Excellence in Quantum Information and Quantum Physics,
University of Science and Technology of China, Hefei, Anhui 230026,
China}
\affiliation{Anhui Province Key Laboratory of Quantum Network,
University of Science and Technology of China, Hefei 230026, China}
\affiliation{Hefei National Laboratory, Hefei 230088, China}

\author{Zi-Jie Chen}
\affiliation{Laboratory of Quantum Information,
University of Science and Technology of China, Hefei 230026, China}
\affiliation{CAS Center For Excellence in Quantum Information and Quantum Physics,
University of Science and Technology of China, Hefei, Anhui 230026,
China}
\affiliation{Anhui Province Key Laboratory of Quantum Network,
University of Science and Technology of China, Hefei 230026, China}
\affiliation{Hefei National Laboratory, Hefei 230088, China}

\author{Yifang Xu}
\affiliation{Center for Quantum Information, Institute for Interdisciplinary Information
Sciences, Tsinghua University, Beijing, China}

\author{Yilong Zhou}
\affiliation{Center for Quantum Information, Institute for Interdisciplinary Information
Sciences, Tsinghua University, Beijing, China}

\author{Hongwei Huang}
\affiliation{Center for Quantum Information, Institute for Interdisciplinary Information
Sciences, Tsinghua University, Beijing, China}

\author{Yunlai Zhu}
\affiliation{Center for Quantum Information, Institute for Interdisciplinary Information
Sciences, Tsinghua University, Beijing, China}

\author{Ziyue Hua}
\affiliation{Center for Quantum Information, Institute for Interdisciplinary Information
Sciences, Tsinghua University, Beijing, China}

\author{Guang-Can Guo}
\affiliation{Laboratory of Quantum Information,
University of Science and Technology of China, Hefei 230026, China}
\affiliation{CAS Center For Excellence in Quantum Information and Quantum Physics,
University of Science and Technology of China, Hefei, Anhui 230026,
China}
\affiliation{Anhui Province Key Laboratory of Quantum Network,
University of Science and Technology of China, Hefei 230026, China}
\affiliation{Hefei National Laboratory, Hefei 230088, China}

\author{Luyan Sun}
\email{luyansun@mail.tsinghua.edu.cn}
\affiliation{Center for Quantum Information, Institute for Interdisciplinary Information Sciences, Tsinghua University, Beijing, China}
\affiliation{Hefei National Laboratory, Hefei 230088, China}

\author{Xiaoting Wang}
\email{xiaoting@uestc.edu.cn}
\affiliation{Institute of Fundamental and Frontier Sciences, University of Electronic Science and Technology of China, Chengdu, 610051, China}

\author{Chang-Ling Zou}
\email{clzou321@ustc.edu.cn}
\affiliation{Laboratory of Quantum Information,
University of Science and Technology of China, Hefei 230026, China}
\affiliation{CAS Center For Excellence in Quantum Information and Quantum Physics,
University of Science and Technology of China, Hefei, Anhui 230026,
China}
\affiliation{Anhui Province Key Laboratory of Quantum Network,
University of Science and Technology of China, Hefei 230026, China}
\affiliation{Hefei National Laboratory, Hefei 230088, China}

%\date{\today}

\begin{abstract}
Deterministic preparation of high-excitation Fock states is a central challenge in bosonic quantum information, with control complexity that generically explodes as the Hilbert space dimension grows. Here we introduce resonant subspace engineering (RSE), a protocol that analytically confines the infinite-dimensional bosonic dynamics to a two-dimensional invariant subspace spanned by an initial coherent state and the target state. State transfer then reduces to a geodesic rotation on a synthetic Bloch sphere, governed by resonance and phase-matching conditions we derive in closed form. For single Fock states, RSE achieves $O(n^{1/4})$ scaling in both evolution time and gate depth, showing a fundamental improvement over existing deterministic schemes. The construction generalizes to $K$-component superpositions via a $(K{+}1)$-dimensional invariant subspace with full $\mathrm{SU}(K{+}1)$ controllability, requiring only 3-5 iterations of operations for superpositions spanning photon numbers 70--100. RSE provides a scalable and analytically transparent framework for large-scale bosonic state engineering and gate synthesis across single- and multimode platforms.
\end{abstract}

\maketitle

\textit{Introduction.-}
Quantum control of bosonic modes is a central topic in quantum optics~\cite{Gerry_Knight_2004, Leonhardt_2010} and bosonic quantum information processing~\cite{RevModPhys.77.513, RevModPhys.84.621}. As the energy eigenstates of bosonic modes, Fock states and their arbitrary superpositions constitute essential non-Gaussian quantum resources, offering broad utility in quantum communication~\cite{Bouwmeester1997}, quantum computation~\cite{PhysRevLett.82.1784, PhysRevResearch.1.033063, PhysRevA.100.032306, Steinbrecher2019, Jin2025}, bosonic quantum error correction~\cite{PhysRevA.64.012310, Waks_2006, PhysRevX.6.031006}, and quantum metrology~\cite{Giovannetti2011,WangNC2019Heisenberg,Deng2024}. In particular, extending Fock state generation into the large photon number regime is not only a pivotal step in advancing bosonic quantum technologies from proof-of-principle experiments toward demonstrations of quantum advantage, but also serves as a powerful probe for exploring novel quantum phenomena at high energy levels.

\begin{figure}
\centering
\includegraphics[width=\columnwidth]{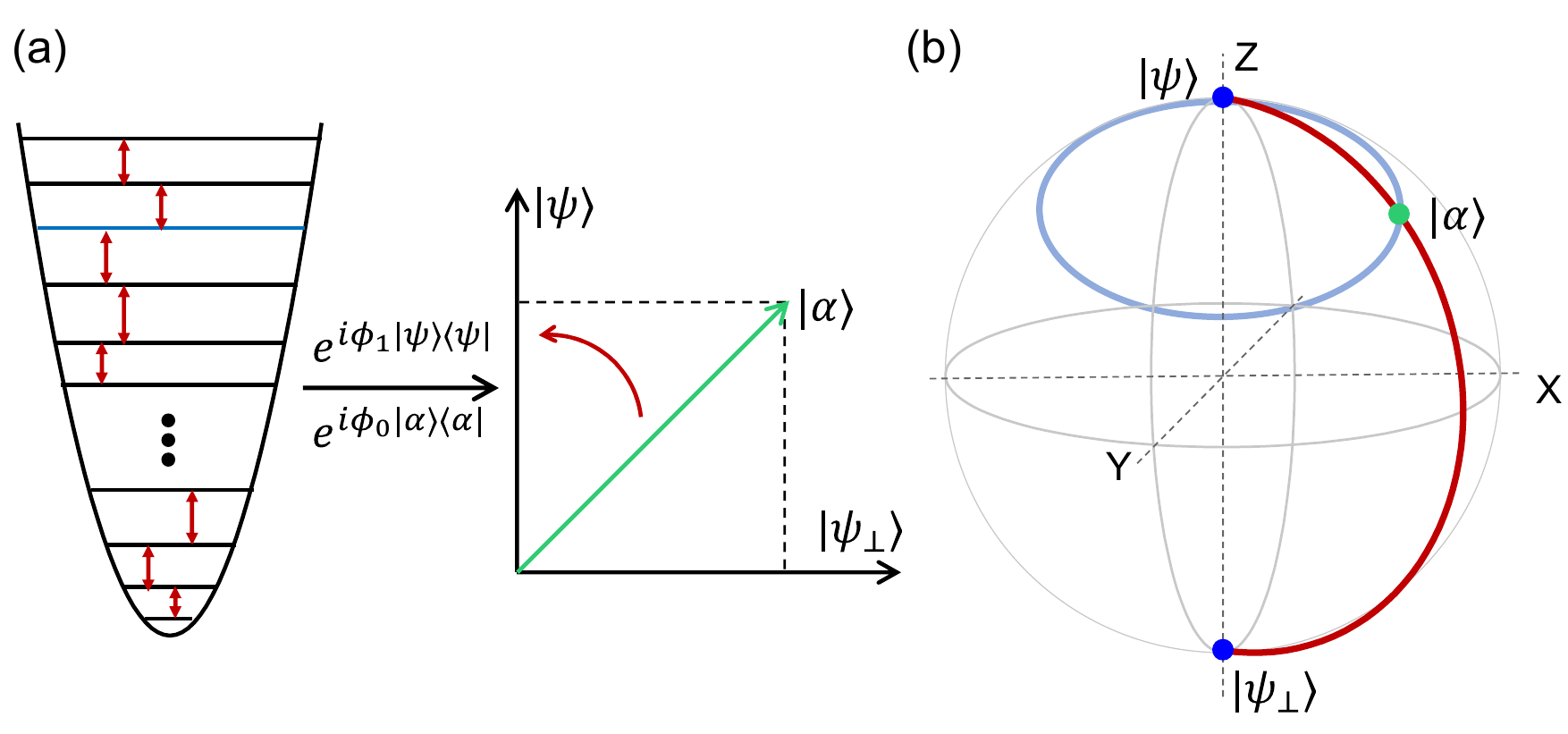}
\caption{Principle of resonant subspace engineering (RSE). (a) Traditional quantum control approaches engineer complex transition paths that connect the initial and target states (left). In contrast, RSE reduces the dynamics to a two-dimensional system spanned by the initial coherent state $|\alpha\rangle$ and the target state $|\psi\rangle$.
(b) Illustration of rotations that carry the initial coherent $|\alpha\rangle$ to the target state on the synthetic Bloch sphere. The resonance condition of the generator Hamiltonian ensures that the quantum state evolves along a geodesic line on the Bloch sphere with the axis in the XY plane.}
\label{Fig1}
\end{figure}

Over the past decades, extensive quantum state engineering schemes have been demonstrated across cavity quantum electrodynamics (QED)~\cite{Varcoe2000, PhysRevLett.76.1055, PhysRevLett.86.3534, PhysRevLett.88.143601, PhysRevLett.125.093603}, superconducting circuits~\cite{Hofheinz2008, Hofheinz2009, Premaratne2017, GU2017}, phononics~\cite{Chu2018}, trapped ions~\cite{PhysRevLett.70.762, PhysRevLett.76.1796}, and optical platforms~\cite{DELLANNO200653, PhysRevLett.115.163603, PhysRevA.100.041802, PhysRevLett.96.213601, PhysRevLett.87.050402, Bimbard2010} to generate Fock states by utilizing the intrinsic bosonic nonlinearity or ancillary atomic nonlinearities. However, advancing the excitations of the number state faces scalability limitations, originating from either the exponentially decreased success probability in measurement-based schemes~\cite{PhysRevLett.108.243602, Sayrin2011, Guerlin2007, PhysRevLett.97.073601, Deleglise2008, PhysRevA.100.041802, PhysRevA.71.021801, Thekkadath2020} or the explosively growing quantum control complexity over the expanding Hilbert space dimension in deterministic protocols.~\cite{PhysRevLett.86.3534, PhysRevLett.115.137002, PhysRevA.92.040303, engelkemeier2021climbingfockladderadvancing}.
Despite recent optics-inspired schemes to reduce the quantum control complexity at the Hamiltonian level~\cite{li2026scalable,xu2026principlesopticsfockspace}, there remains no clear mechanism that systematically reduces the quantum control complexity at the Hilbert space level across the whole process of quantum state evolution, while also meeting the demand of analytical solvability, high control efficiency, and experimental feasibility. The difficulty becomes even more pronounced for high-excitation, multi-component superpositions: while universal bosonic control tools have been established~\cite{PhysRevLett.76.1055, PhysRevLett.115.137002, PhysRevA.92.040303}, preparing a specific large-$n$ superposition typically relies on nontrivial compilation or numerically optimized control, and the experimental overhead tends to grow quickly as the mean photon number and the number of addressed components increase.

In this Letter, we introduce resonant subspace engineering (RSE), a protocol that reduces the state-preparation dynamics of an infinite-dimensional bosonic mode to an engineered two-dimensional subspace spanned by the initial and target states. We identify the Hamiltonian to realize Pauli rotations in this subspace, thereby ensuring perfect state transfer from the initial state to a large Fock state or an arbitrary Fock-state superposition. To ensure experimental feasibility, we compile the continuous evolution into a discrete gate sequence using only displacement and selective number-dependent arbitrary phase (SNAP) gates~\cite{PhysRevLett.115.137002}. Starting from an initial coherent state, both the evolution time and circuit depth for preparing $\ket{n}$ scale as $\mathcal{O}(n^{1/4})$. Our RSE provides a general design principle for controlling quantum dynamics in high-dimensional bosonic quantum systems.

\textit{Principle of RSE.-} Before discussing the practical implementation of arbitrary Fock state preparation, we propose the RSE framework to realize efficient conversion between a set of target states in an infinite-dimensional Hilbert space. Figure~\ref{Fig1}(a) illustrates the key idea of RSE. In conventional optimal quantum control approaches, e.g., GRAPE~\cite{UControlKhaneja2005JMRGRAPE, ChenZJ2025SciAdvOpenGRAPE} and SNAP, in the infinite-dimensional Hilbert space of a bosonic mode, we need to engineer all potential transition paths that connect the initial and target states, which are numerically challenging and computationally resource-demanding for highly excited Fock states. Instead, we propose an idea inspired by the amplitude amplification in Grover's search method for qubit systems~\cite{10.1145/237814.237866, PhysRevLett.113.210501, 595153, PhysRevLett.80.4329, 3fzf-wsr2}, to reduce the Hilbert space to two dimensions (We show in the Discussion that RSE extends naturally to multi-level subspaces).

Suppose we can implement the generalized oracle unitary operations (GOOs) $O(\ket{\alpha},\phi_0)=e^{-i\phi_0 |\alpha\rangle\langle\alpha|}$ and
$O(\ket{\psi},\phi_1)=e^{-i\phi_1 |\psi\rangle\langle\psi|}$,
which imprint tunable phases on the target state $|\psi\rangle$ and a reference state $|\alpha\rangle$, respectively.
When these two oracles are applied alternately in small steps, the Trotter-Suzuki decomposition~\cite{Suzuki1976,doi:10.1126/science.273.5278.1073} shows that the stroboscopic dynamics are governed by an effective Hamiltonian
\begin{equation}\label{eq:Heff}
  H \propto \phi_1\,|\psi\rangle\langle\psi| + \phi_0\,|\alpha\rangle\langle\alpha|.
\end{equation}
Despite operating on an infinite-dimensional bosonic Hilbert space, $H$ acts nontrivially only on the two-dimensional subspace $\mathcal{S}=\mathrm{span}\{|\psi\rangle,|\alpha\rangle\}$, since both $H|\psi\rangle$ and $H|\alpha\rangle$ lie within $\mathcal{S}$.
The entire infinite-dimensional dynamics is thereby \emph{exactly} reduced to a two-level problem, regardless of the photon numbers involved. This reduction is the central insight of RSE and has a transparent geometric picture.  We introduce the orthogonal basis $\{|\psi\rangle,|\psi_\perp\rangle\}$ in $\mathcal{S}$, with
\begin{equation}\label{psi_perp}
  |\psi_\perp\rangle = \frac{|\alpha\rangle - \mu\,|\psi\rangle}{\sqrt{1-|\mu|^2}},\qquad
  \mu \equiv |\mu|\,e^{i\theta} \equiv \langle\psi|\alpha\rangle.
\end{equation}
Projecting $H$ onto $\mathcal{S}$ yields a $2\times 2$ matrix $H_\mathcal{S}$, whose diagonal elements (the effective detuning) and off-diagonal elements (the effective coupling) are fully determined by $\phi_0$, $\phi_1$, and the overlap $\mu$.
This matrix represents a general rotation generator on a synthetic Bloch sphere spanned by $|\psi\rangle$ and $|\psi_\perp\rangle$ [Fig.~\ref{Fig1}(b)], providing an analytically transparent SU(2) control framework in the subspace $\mathcal{S}$ while the complementary space $\mathcal{S}^\perp$ remains completely dark. Therefore, RSE is fundamentally a \emph{gate} construction. By composing sequences of $O(\ket{\psi},\phi_1^{(j)})$ and $O(\ket{\alpha},\phi_0^{(j)})$ with iteration-dependent parameters $\{\phi_0^{(j)},\phi_1^{(j)}\}$, one can synthesize any target SU(2) gate within $\mathcal{S}$.

In particular, Grover's search realizes the state mapping from $\ket{\alpha}$ to $\ket{\psi}$ by setting $\phi_0=\phi_1=\pi$. RSE achieves higher efficiency by engineering the state evolution to follow the geodesic, which is the shortest path on the Bloch sphere [Fig~\ref{Fig1}(b), red line], and we propose the two conditions on the Hamiltonian that generates the rotation:

(i) \emph{Resonance condition.}
Writing $H$ in the basis of $\{\ket{\psi},\ket{\psi_\perp}\}$ gives
\begin{equation}
    H_{\mathcal{S}}=
    \begin{pmatrix}
        \bra{\psi}H\ket{\psi} & \bra{\psi}H\ket{\psi_\perp}\\[2pt]
        \bra{\psi_\perp}H\ket{\psi} & \bra{\psi_\perp}H\ket{\psi_\perp}
    \end{pmatrix}.
    \label{HS}
\end{equation}
We require the two levels to be on resonance,
\begin{equation}
    H_{11}=H_{22},
    \label{resonance}
\end{equation}
which removes the detuning (the $\sigma_z$ component) in $H_{\mathcal S}$ and leaves a purely transverse coupling set by $H_{12}$, corresponding to a geodesic rotation (axis in the XY plane) between $\ket{\psi}$ and $\ket{\psi_\perp}$. Here $H_{ij}$ denotes the matrix elements of $H_{\mathcal{S}}$.

(ii) \emph{Phase-matching condition.}
The XY-plane geodesic rotation must also carry the initial state $\ket{\alpha}$; equivalently, the coupling phase of $H_{12}$ must match the relative phase $\theta$ in Eq.~\eqref{psi_perp}. We therefore require $\arg(H_{12})=\theta+\frac{\pi}{2}\quad (\mathrm{mod}\ 2\pi)$
so that the induced resonant rotation can steer $\ket{\alpha}$ precisely to $\ket{\psi}$ at time $T$.

Under the two conditions, $H_{\mathcal S}$ is proportional (up to an overall phase) to the Pauli operators in $\mathcal S$, yielding the deterministic transfer time
\begin{equation}
    T = \frac{\arccos(|\mu|)}{|H_{12}|}.
    \label{conversion_time}
\end{equation}
For a more general case that the phase of $H_{12}$ cannot be easily tuned, an additional rotation with axis nonparallel to the XY plane is also necessary to carry the initial state $\ket{\alpha}$ to the trajectory of the geodesic rotation. The combination of the XY-plane rotation and off-plane rotation is sufficient to realize a universal SU(2) transformation on the sphere.

\textit{Physical implementation of GOOs.-} The GOOs required by RSE arise naturally in circuit QED systems~\cite{CAI202150,MA20211789}, where a bosonic cavity mode is coupled to an ancillary qubit, a model that can also be generalized to optical cavity QED systems. In the dispersive regime, the cavity-qubit interaction takes the form $H_{\rm disp} = -\chi\, a^\dag a\,|e\rangle\langle e|$, where $\chi$ is the dispersive coupling strength and $|e\rangle$ denotes the excited state of the qubit with transition frequency $\omega_\mathrm{q}$. This interaction renders the qubit transition frequency dependent on the photon number $n$, enabling \emph{photon-number-resolved} control: by frequency-selectively driving on the qubit ($\omega_\mathrm{q}-n\chi$ for Fock state $|n\rangle$), one can imprint an arbitrary phase $\theta_n$ on each individual Fock component independently~\cite{PhysRevA.92.040303,PhysRevLett.115.137002}, i.e. implementing unitary
$U=\sum_n e^{i\theta_n}|n\rangle\langle n|$. This capability provides a direct physical realization of the GOO on Fock states or their combinations. Additionally, combining the GOO $O(\ket{0},\phi_0)$ and the displacement operation $D(\alpha)$ with amplitude $\alpha$, we can implement the GOO for a coherent state $|\alpha\rangle_\mathrm{c}$ through
\begin{equation}\label{eq:Oa}
  O(|\alpha\rangle_\mathrm{c},\phi_0) = e^{-i\phi_0|\alpha\rangle_\mathrm{c}\langle\alpha|_\mathrm{c}}.
\end{equation}
Then, the RSE protocol is implemented by alternating applications of $O(|\alpha\rangle_\mathrm{c},\phi_0)$ and the GOO for the target Fock state $O(|n\rangle,\phi_1)$  using only displacement pulses and SNAP gates. The state-preparation sequence is deterministic and can be constructed analytically, providing a powerful and convenient approach for manipulating the infinitely large Hilbert space of harmonic oscillators.

\textit{RSE for a single Fock state.-}
We first specialize the RSE framework to the preparation of a single Fock state $\ket{\psi}=\ket{n}$ by applying GOOs for a coherent state ($|\alpha\rangle_\mathrm{c}$) and the target Fock state, which leads to the effective Hamiltonian of a two–level subspace as
\begin{equation}
    H_n = \omega_n|n\rangle\langle n| + \Omega|\alpha\rangle_\mathrm{c}\langle\alpha|_\mathrm{c},
    \label{Ham_single}
\end{equation}
where $\omega_n$ and $\Omega$ are the effective frequency shifts acting on the Fock and coherent states, respectively. According to Eq.~(\ref{psi_perp}), we have $\mu=\alpha_n\equiv\langle n|\alpha\rangle=e^{-|\alpha|^2/2}\alpha^n/\sqrt{n!}$  for a Poisson distribution of Fock state coefficients and can derive the corresponding $\ket{n_\perp}$, so that $\mathcal S=\mathrm{span}\{\ket{n},\ket{n_\perp}\}$. The resonance condition of the Hamiltonian $H_n$ requires
\begin{equation}
    \omega_n = \Omega(1-2|\alpha_n|^2).
    \label{Fock-resonance}
\end{equation}
To enforce the phase-matching condition, we apply $U_n=e^{-i\pi/2\ket{n}\bra{n} }$ prior to the resonant evolution, corresponding to a Pauli-$Z$ rotation that aligns $|\alpha\rangle$ with the rotation trajectory of $H_n$.

\begin{figure}[t]
  \centering
  \includegraphics[width=1.0\linewidth, ,trim=0 8 0 0,clip]{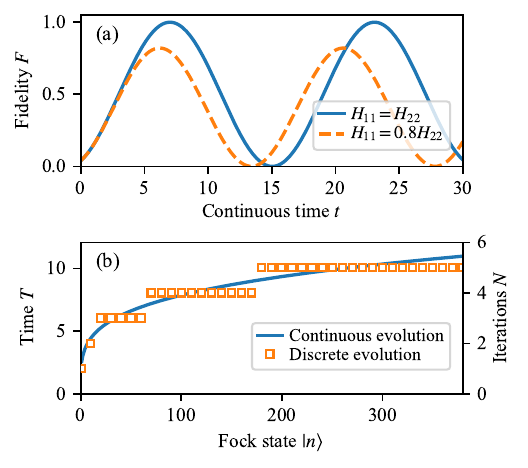}
  \caption{(a) Fidelity dynamics for preparing $\ket{n=100}$ from $\ket{\alpha=10}$. The blue solid (orange dashed) curve shows the evolution for matched $H_{11}=H_{22}$ (mismatched $H_{11}=0.8H_{22}$). (b) Required evolution time $T$ (blue solid, left axis) and optimized iteration number $N$ (orange markers, right axis) versus target photon number $n$ for a coherent input $\ket{\alpha=\sqrt{n}}$, illustrating the sublinear $\mathcal{O}(n^{1/4})$ scaling.}
  \label{Fig2}
\end{figure}

Then, the dynamics in $\mathcal{S}$ reduce to a resonant Rabi oscillation between $\ket{n}$ and $\ket{n_\perp}$ that passes through $U_n\ket{\alpha}$, with the  required evolution time [Eq.~\eqref{conversion_time}] being
\begin{equation}
    T
    =\frac{\arccos(|\alpha_n|)}{\Omega|\alpha_n|\sqrt{1-|\alpha_n|^2}}
    \le \frac{\pi}{2\Omega|\alpha_n|}=\mathcal{O}(n^{1/4}).
    \label{single_time}
\end{equation}
To verify this scheme, we numerically simulate the preparation of $\ket{n=100}$ by RSE with GOOs on a coherent state and the target state. Since the photon-number distribution of a coherent state is Poissonian, we choose $\alpha = \sqrt{n}$ to maximize the overlap $|\alpha_n|^2$ and thus minimize $T$. As shown in Fig.~\ref{Fig2}(a) (blue curve), for $\omega_n = 1-2\alpha_n^2$ and $\Omega=1$, the unitary evolution $e^{-iH_n t}$ drives the system to the target state at the predicted time $T \approx 7.8677$, followed by periodic oscillations between $\ket{\alpha}_\mathrm{c}$ and the target state $\ket{n}$. In contrast, for a mismatched choice (e.g., $H_{11} = 0.8H_{22}$, orange dashed curve), the evolution fails to reach high fidelity, showing a faster oscillation behavior that features a driven two-level atom with finite detuning. The results highlight the necessity of subspace resonance for successful state transfer.

It should be noted that combinations of SNAP gates and displacement have also been demonstrated to prepare arbitrary Fock state superpositions~\cite{PhysRevLett.115.137002, PhysRevA.92.040303}. The significance of RSE protocol is that it reduces the state dynamics to a two-dimensional subspace and gives the conditions on the alternating phases defined by Eq.~\eqref{Fock-resonance}, which greatly reduces the quantum control complexity and ensures Fock state preparation with short time and low circuit depth.
Figure~\ref{Fig2}(b) shows the relationship between the Fock number and the required evolution time,
which is sublinear in $n$ and is consistent with the continuous-time scaling in Eq.~\eqref{single_time}.

Beyond this analytic angle choice, we note that GOOs generate rotations about two non-parallel axes in the same two-dimensional subspace $\mathcal S$, enabling universal SU(2) control within $\mathcal S$. In practice, one can therefore directly optimize the step angles $\{b_j,c_j\}$ in each iteration $O(|\alpha\rangle_\mathrm{c},c_j)O(|n\rangle,b_j)$ to maximize the final fidelity. The optimized discrete iteration counts in Fig.~\ref{Fig2}(b) (orange markers) follow the same sublinear trend. In particular, preparing $\ket{n=100}$ and $\ket{n=380}$ requires only $N=4$ and $N=5$ iterations of GOOs, respectively, significantly outperforming traditional methods and demonstrating the practical efficiency of the RSE protocol for large-$n$ Fock-state preparation.

\begin{figure}[t]
\centering
\includegraphics[width=0.95\linewidth,trim=0 0 0 0,clip]{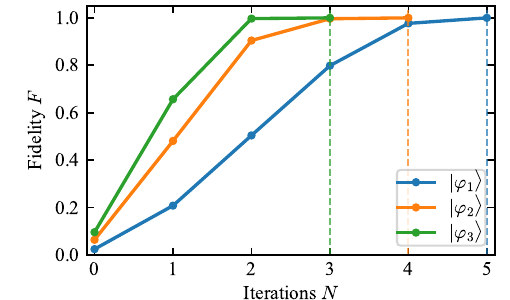}
\caption{Preparation of Fock-state superpositions from $\ket{\alpha=\sqrt{88}}$ with optimized generalized oracle operation angles. Target states: $\ket{\varphi_1}=\sqrt{0.3}\ket{70}+\sqrt{0.7}\ket{100}$, $\ket{\varphi_2}=\sqrt{0.2}\ket{70}+\sqrt{0.5}\ket{85}+\sqrt{0.3}\ket{100}$, and $\ket{\varphi_3}=\sqrt{0.1}\ket{70}+\sqrt{0.3}\ket{80}+\sqrt{0.4}\ket{90}+\sqrt{0.2}\ket{100}$. $F=1$ is obtained with $N=5, 4, 3$ iterations, respectively.}
  \label{Fig3}
\end{figure}

\textit{RSE in higher dimensions.-}
The deterministic preparation of arbitrary Fock superpositions, i.e., target states of the form $|\psi\rangle=\sum_{n\in A_M} c_n|n\rangle$ with $A_M$ denoting the set of $M$ selected Fock states, is required for general applications. However, we note that simultaneously driving multiple tones on the qubits $\omega_\mathrm{q}-n\chi$ ($n\in A_M$) do not directly generate the GOO for $|\psi\rangle$. Instead, such process extends the RSE principle beyond two dimensions. In a more general setting, suppose we introduce a set of GOOs $O({\ket{\psi_j}},\phi_j)=e^{-i\phi_j |\psi_j\rangle\langle\psi_j|}$ ($j=1,2,...,K$) with $\bra{\psi_j}\psi_k\rangle=\delta_{jk}$ being orthogonal to each other, associated with a reference state $|\alpha\rangle$. The effective Hamiltonian becomes
\begin{equation}
    H\propto\phi_0\,|\alpha\rangle\langle\alpha|+\sum_{j=1}^K \phi_j\,|\psi_j\rangle\langle\psi_j|,
\end{equation}
which is confined to the $(K+1)$-dimensional invariant subspace $\mathcal{S}_K = \mathrm{span}\{|\psi_1\rangle,|\psi_2\rangle,...,|\psi_K\rangle,|\psi_\perp\rangle\}$, where $|\psi_\perp\rangle$ is the component of $|\alpha\rangle$ orthogonal to all $|\psi_j\rangle$ ($  j\in{1,2,...,K}$). Consequently, the capability of arbitrary gate operations and state preparations in the two-dimension ($K=1$), as demonstrated above, can be directly extended to arbitrary $K$ as long as $\langle\alpha|\psi_j\rangle\neq0$ for all $|\psi_j\rangle$. That is, the GOOs for $K$ mutually orthogonal target states together with a reference state generate a $(K+1)$-dimensional subspace with full $\mathrm{SU}(K+1)$ control capability, and similar  RSE conditions are applicable.

As examples, the GOOs on individual Fock states satisfying mutual orthogonality condition enable such high-dimensional RSE. We numerically simulate the preparation of three representative target superpositions $\ket{\varphi_1}$, $\ket{\varphi_2}$, and $\ket{\varphi_3}$ from $\ket{\alpha=\sqrt{88}}$, as shown in Fig.~\ref{Fig3}. We directly optimize the sequence parameters in GOOs $O(\ket{\alpha},c_j)$ and $O(\ket{\psi_j},\phi_j)$ to maximize the final fidelity. The results show that $\ket{\varphi_1}$, $\ket{\varphi_2}$, and $\ket{\varphi_3}$ can be prepared with only $N=5$, $4$, and $3$ iterations, respectively, demonstrating the practical efficiency of the RSE approach for arbitrary superpositions of large Fock states.

\textit{Discussion.-}
Beyond Fock-state preparation, the RSE mechanism directly enables logical gate operations on bosonic error-correcting codes whose codewords are built from coherent-state superpositions. Consider the cat codes, whose logical states are superpositions of basis coherent states $\ket{e^{ik\pi/d}\alpha}_\mathrm{c}$ ($k=0,1,...,2d-1$, and $d\in\mathbb{N}$)~\cite{CAI202150,Li2017}. For large $\alpha$, the overlaps among these basis coherent states are exponentially suppressed, making them approximately orthogonal. The high-dimensional RSE then applies directly by alternately applying the GOOs for $2d$ basis coherent states and a reference Fock state, and arbitrary logical states and the logical gate on the logical qubit (qudit) can be implemented in a $(2d+1)$-dimensional invariant subspace.  In contrast to conventional approaches that require numerical optimization over a large Hilbert space and re-optimization whenever $\alpha$ is changed, RSE yields analytically determined control parameters for all $\alpha$.

A particularly powerful generalization of RSE arises in multimode architectures, as a single transmon qubit can be dispersively coupled to multiple microwave cavity modes simultaneously. Recently, the selective operation on the joint Fock states across up to five modes has been demonstrated~\cite{Chakram2022}, laying the ground for the multiple-mode SNAP operations. Combining with the single-mode coherent state GOOs for each mode and the GOOs for joint Fock states, it is possible to engineering the non-trivial entangled multimode quantum states. The exploration of the invariant subspace across multiple modes will be of fundamental interest, enabling deterministic preparation of entangled multimode states such as NOON states ($(\ket{n,0} + \ket{0,n})/\sqrt{2}$) and the multimode cat states. This extension opens a pathway toward deterministic generation of large-scale multimode entangled bosonic states, which constitute key resources for multimode bosonic error correction, distributed quantum sensing, and continuous-variable quantum networks.

\textit{Conclusion.-} In summary, we have proposed a resonant-subspace engineering mechanism to deterministically synthesize arbitrary Fock-state superpositions. By confining the dynamics to a two-dimensional invariant subspace and enforcing closure, resonance, and phase-matching conditions, we construct effective single-qubit-like rotations that implement direct state transfer from an initial state to target Fock superpositions. The resulting continuous evolution can be compiled into experimentally accessible displacement and SNAP operations. Starting from coherent inputs, our protocol prepares Fock states $|n\rangle$ with evolution time and iteration number scaling as $\mathcal{O}(n^{1/4})$, and the same subspace construction extends naturally to arbitrary superpositions, including those with large photon-number separation. The protocol is fully constructive and experimentally compatible, relying only on displacement and SNAP gates, thereby providing a scalable approach to generating nonclassical bosonic resource states for continuous-variable quantum computing~\cite{RevModPhys.77.513, RevModPhys.84.621, PhysRevLett.82.1784}, photonic state engineering~\cite{DELLANNO200653, PhysRevLett.115.163603, PhysRevA.100.041802, PhysRevLett.96.213601, PhysRevLett.87.050402, Bimbard2010}, and quantum metrology~\cite{Giovannetti2011, Deng2024}.

\smallskip{}
\begin{acknowledgments}
This work was funded by the National Natural Science Foundation of China (Grants No. 92265210, 12550006, 92365301, 92565301, 92165209, 12547179, 12574539, and 92265208) and the Quantum Science and Technology-National Science and Technology Major Project (2021ZD0300200). X.W. also acknowledge the support from the Sichuan Science and Technology Program (2025YFHZ0336). This work is also supported by the Fundamental Research Funds for the Central Universities, the USTC Research Funds of the Double First-Class Initiative, the supercomputing system in the Supercomputing Center of USTC the USTC Center for Micro and Nanoscale Research and Fabrication.
\end{acknowledgments}

\end{document}